\documentclass[%
 twocolumn,
 amsmath,amssymb,
 aps,
 prl,
]{revtex4-2}

\usepackage{graphicx}
\usepackage{dcolumn}
\usepackage{bm}
\usepackage{hyperref}
\usepackage[mathlines]{lineno}
\usepackage{lipsum}
\usepackage{xcolor}

\begin{document}

\preprint{APS/123-QED}

\title{The near-wall cycle for skin-friction generation revealed through explainable deep learning}

\author{Andrés Cremades}
\affiliation{Instituto Universitario de Matemática Pura y Aplicada, 
Universitat Politècnica de València, Valencia, 46022, Spain}

\author{Sergio Hoyas}
\affiliation{Instituto Universitario de Matemática Pura y Aplicada, 
Universitat Politècnica de València, Valencia, 46022, Spain}

\author{Ricardo Vinuesa}
\affiliation{Department of Aerospace Engineering, \\
University of Michigan, Ann Arbor, MI 48109, United States}

\date{\today}

\begin{abstract}
Skin friction in wall-bounded turbulence is produced by intermittent near-wall motions, yet conventional coherent-structure definitions do not identify which individual events generate wall-shear stress, and thus friction. We train neural networks to predict the future velocity field and wall-shear-stress distribution in turbulent channel flow, and use SHAP-derived importance maps to identify the the input regions most influential for each prediction. The dominant events appear as paired objects: an upstream velocity-relevant region is associated with high-momentum motion toward the wall, while a downstream friction-relevant region marks the enhanced wall-shear-stress left in its wake. Tracking these pairs reveals a recurrent cycle of growth, streamwise elongation, decay, and occasional splitting into new wall-shear-producing events. These results redefine near-wall coherent structures not by what the flow looks like, but by what they do to wall-shear stress.

\end{abstract}
\maketitle

Turbulence remains one of the major open problems in physics, while also being one of the phenomena with the widest range of practical applications~\cite{sreenivasan2025turbulence}. In wall-bounded flows, one of its most important consequences is skin friction: the turbulent transfer of momentum toward the wall produces wall-shear stress, which is responsible for a large fraction of drag in many engineering systems. Understanding which flow structures are responsible for this wall-shear generation is therefore both a fundamental and practical problem.

Historically, two complementary viewpoints have shaped our understanding of turbulent flows. The first is statistical: beginning with the Reynolds-averaged equations~\cite{rey83}, turbulence has been described in terms of mean quantities and correlations, leading to the closure problem and motivating scaling theories, numerical simulations, turbulence models, and, more recently, data-driven approaches~\cite{kolmogorov1941,frisch1995turbulence,smagorinsky1963,kim1987,pope2001turbulent,cremades2025}. 

The second viewpoint is structural: turbulence is interpreted in terms of coherent motions, i.e., organized flow regions that can be distinguished from the turbulent background and play a relevant role in the transfer of momentum and energy~\citep{Jimenez2018}. In wall-bounded turbulence, this structural viewpoint is particularly relevant because near-wall motions sustain turbulence and regulate the momentum transfer that ultimately produces skin friction. 

Several families of coherent structures have therefore been used to describe the organization of the near-wall region. Streaks represent elongated low- and high-speed regions generated by the lift-up mechanism~\cite{kline1967structure}, quadrant events identify intense Reynolds-stress-producing motions such as ejections and sweeps~\citep{wallace1972}, vortices are commonly defined from local properties of the velocity-gradient tensor~\citep{jeong1995}, and hairpin-like structures provide a geometrical picture of wall turbulence dating back to the model of~\citet{theodorsen1952} and later boundary-layer studies~\citep{head1981,perry1982,adrian2000}.

These descriptions have been essential for interpreting the near-wall cycle~\cite{Jimenez2018}. However, they all rely on an a priori criterion to identify coherent structures, such as velocity intensity, Reynolds-stress contribution, kinematic indicators, or prescribed morphology. As a result, they select flow regions that satisfy a predefined criterion, rather than structures selected by their relevance to wall-shear generation. This limitation is particularly evident in shape-based descriptions, since the experimental identification of hairpin vortices at high Reynolds numbers remains elusive~\citep{dennis2011}.

\begin{figure*}[t!]
    \centering
    \includegraphics[width=1\linewidth]{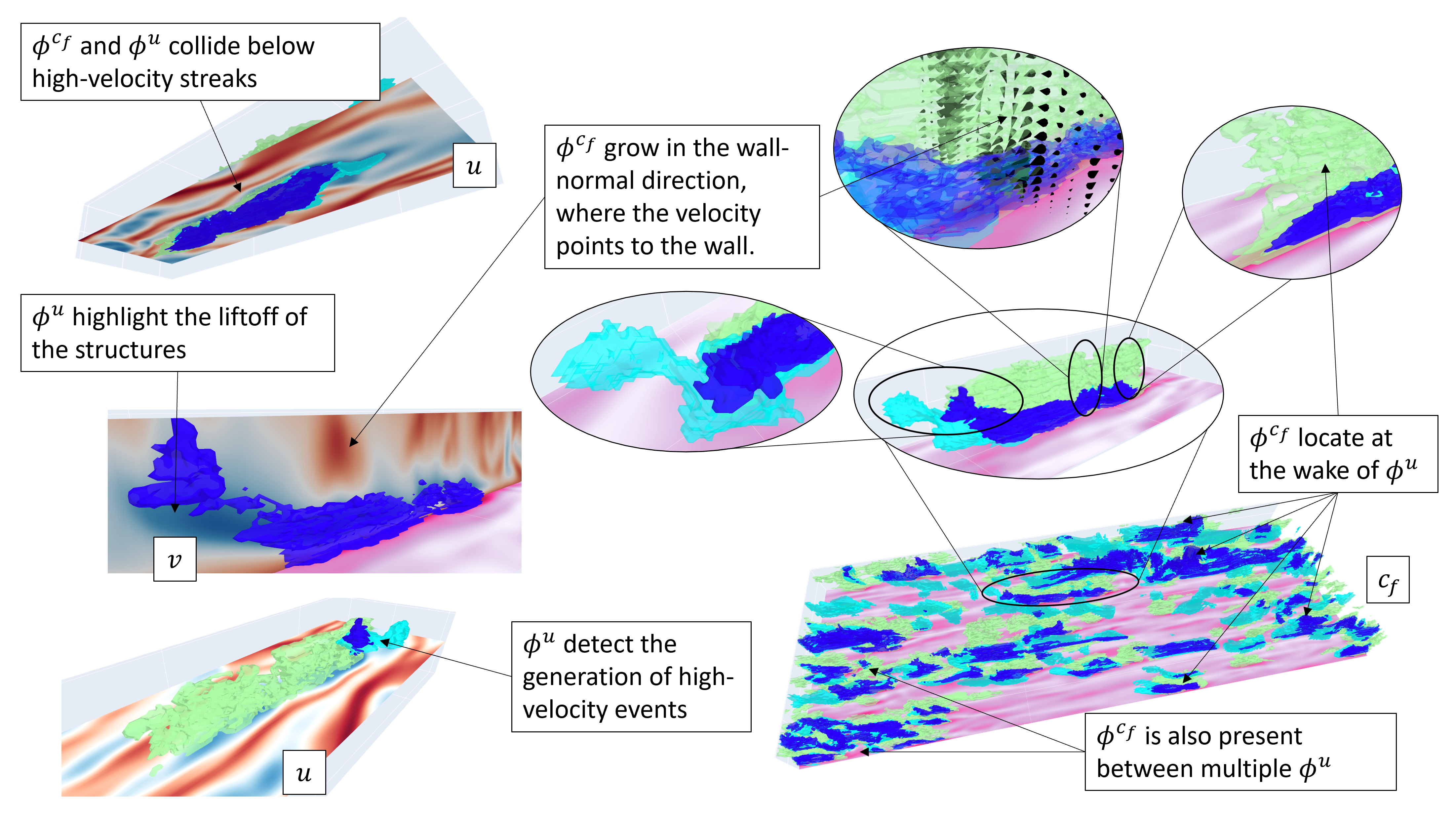}
    \caption{Instantaneous visualization of the VS pairs. Velocity-based SHAP structures are shown in light blue, wall-shear-based SHAP structures in light green, and their overlap in dark blue. Wall-shear stress is represented by the pink colormap, with darker colors indicating higher wall-shear stress, and velocity fluctuations are shown with a blue-to-red colormap, indicating (blue) high and (red) low velocity. The component of the velocity presented in each panel is indicated in the respective text box.}
    \label{fig:view_pairs_shap}
\end{figure*}

In the present work, we adopt a data-driven version of this structural viewpoint. Rather than prescribing any criteria, we ask which regions of the flow are most relevant to the prediction of wall-shear stress. To this end, we train neural-network models to predict the velocity field and the wall-shear distribution, and use explainable artificial intelligence to identify the input regions most important to each prediction.

Briefly, the database analyzed in this work is obtained from a direct numerical simulation of turbulent channel flow of height $2h$ at $Re_\tau=u_\tau h/\nu=125$, where $u_\tau$ is the friction velocity and $\nu$ is the kinematic viscosity. Further details of the simulation are provided in the Appendix. The computational domain has dimensions $8\pi h \times 2h \times 3\pi h$, with periodic boundary conditions in the streamwise and spanwise directions and no-slip boundary conditions at $y/h=\pm 1$. Wall-normal distances from the nearest wall are also reported in viscous units as $y^+=(h-|y|)u_\tau/\nu$, which measures the distance to the wall normalized by the viscous length scale.

This domain is sufficiently large to contain the near-wall structures relevant to the present analysis at this Reynolds number. Since the focus of this work is the near-wall region, this Reynolds number captures the self-sustaining near-wall cycle and the associated wall-shear-producing motions~\cite{hamilton1995regeneration,jimenez1999autonomous}. Moreover, the near-wall dynamics are known to exhibit a strong degree of universality in viscous units, with small-scale energy, production, dissipation, and transport showing weak Reynolds-number dependence when expressed in inner scaling~\cite{leemoser2019}.

The three-dimensional velocity field was stored during the simulation at intervals of $\Delta t^+ = 5$ viscous time units, where $t^+=tu_\tau^2/\nu$. In total, the database contains 18,000 three-dimensional fields, which are used for model training and testing and, subsequently, for explainability analysis leading to the identification of friction-relevant structures.

The velocity fields are used to train two U-Net models. The first model, introduced in~\citet{cremades2025classically}, predicts the future velocity field. The second model predicts the local skin-friction coefficient, $c_f(x,z,t)$, which is directly proportional to the local wall-shear stress. In both cases, the input velocity field at time $t^+$ is mapped to the corresponding target at $t^+ + \Delta t^+$. We then compute SHAP values using the gradient-SHAP algorithm~\cite{erion2021}. Further details on the U-Net architecture and the gradient-SHAP algorithm are provided in the Appendix.

The first attribution field, $\phi^{u}$, quantifies the contribution of each grid point to the temporal evolution of the flow, identifying the velocity features most relevant for reconstructing the future velocity field. The second attribution field, $\phi^{c_f}$, measures the contribution of the three-dimensional velocity field to the reconstruction of the two-dimensional skin-friction field. This second model constitutes the central methodological extension with respect to prior work: while velocity-based SHAP structures have been analyzed previously, the skin-friction-prediction model enables a spatially resolved identification of the flow regions most relevant to local wall-shear generation, without relying on indirect proxies such as Reynolds-stress events, vortical indicators, or prescribed geometrical criteria.

\begin{figure*}[t!]
    \centering
    \includegraphics[width=0.95\linewidth]{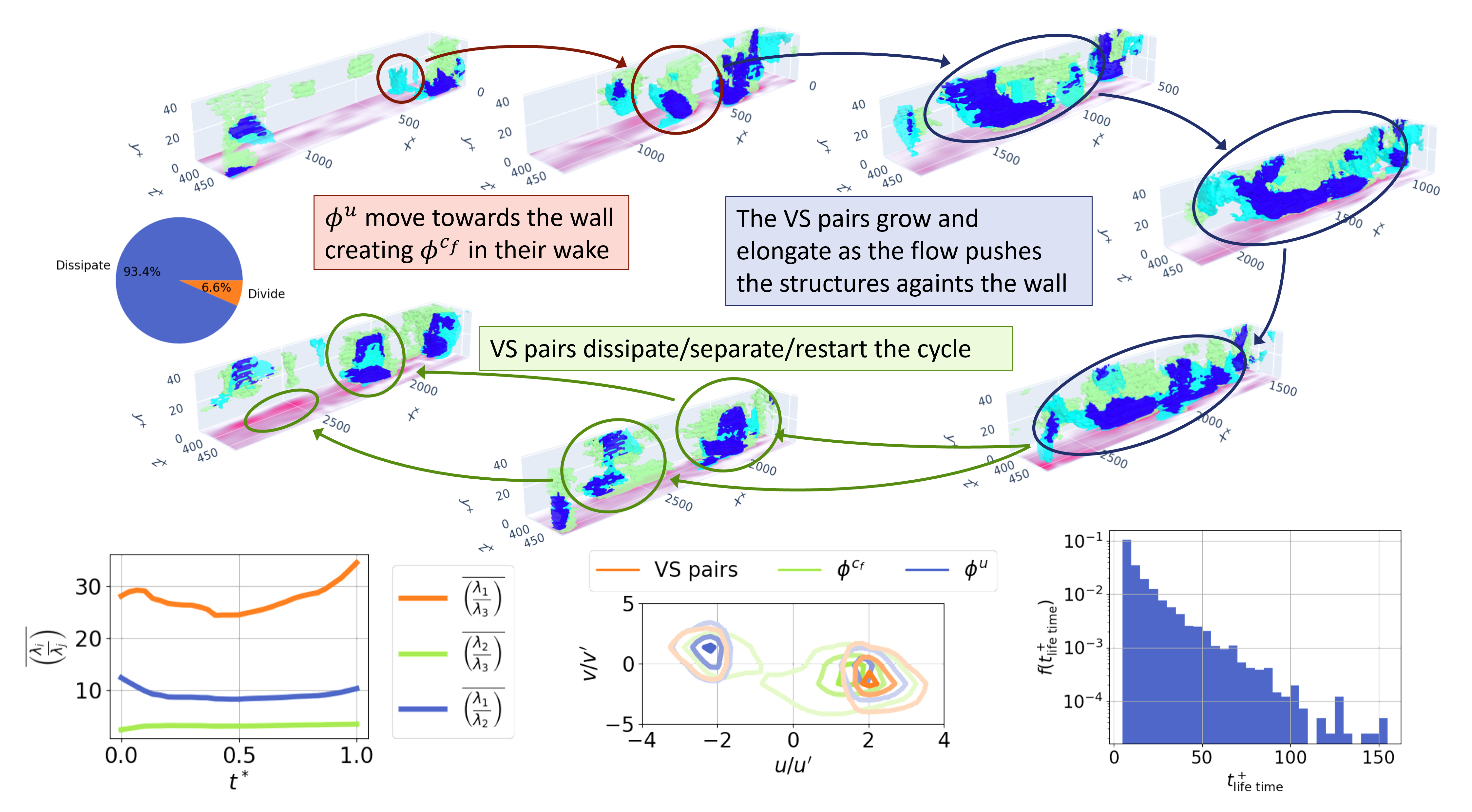}
    \caption{Schematic representation of the VS cycle. The top figure shows a visualization of the friction cycle and the percentage of structures that divide, while the bottom row presents the cycle statistics (from left to right): aspect ratio, velocity distribution inside the structures, and lifetime of the structures. Note that the VS cycle is presented for $y^+<40$.}
    \label{fig:cf_cycle_shap}
\end{figure*}

\begin{figure}
    \centering
    \includegraphics[width=0.99\linewidth]{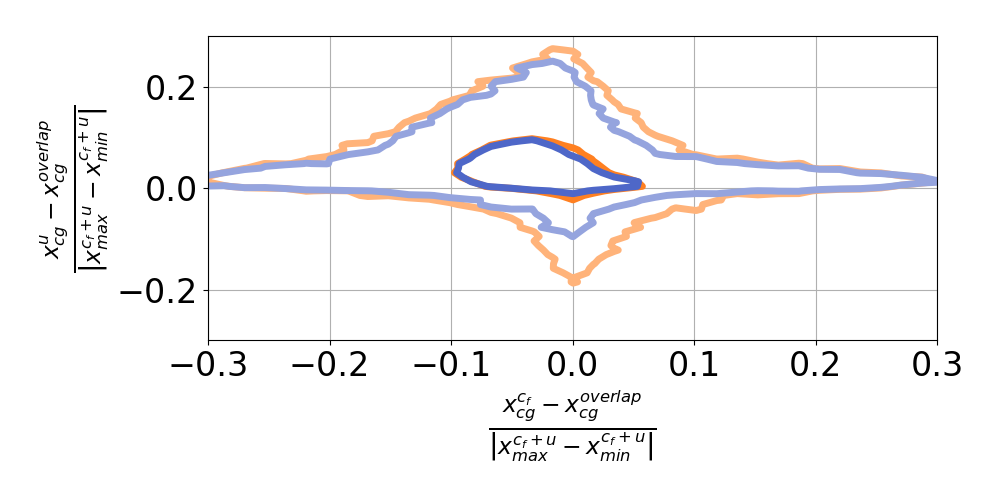}

    \caption{Normalized joint probability density function of the relative location of the centers of gravity of the VS pairs from the wall to $y^+\approx 40$. Orange: distribution based on the centers of gravity of the structures. Blue: distribution weighted by the total SHAP value for wall-shear reconstruction in the overlapping region, divided by the volume of the overlap. Dark contour $10\%$ of the maximum, light contour $1\%$ of the maximum.}
    \label{fig:cg_pairs_shap}
\end{figure}

For each model, SHAP values are used to identify the regions of the input velocity field that are most relevant to the corresponding prediction. This allows us to define a new class of paired structures, referred to as velocity-shear-coherent pairs (VS pairs). Each VS pair consists of a velocity-relevant component, extracted from $\phi^{u}$, and a skin-friction-relevant component, extracted from $\phi^{c_f}$. As shown in \autoref{fig:view_pairs_shap}, the velocity-relevant component is typically located upstream, while the skin-friction-relevant component appears downstream as a localized signature of enhanced wall shear. Technical details on the SHAP computation, thresholding, pairing criterion, and percolation-based extraction of individual structures are provided in the Appendix.

From a classical structural viewpoint, the near-wall cycle is commonly interpreted as a self-sustaining process that organizes the dynamics of the buffer layer~\cite{jimenez1999autonomous,hamilton1995regeneration,waleffe1997self}. In this picture, quasi-streamwise vortices redistribute mean momentum through the lift-up mechanism, giving rise to elongated low- and high-speed streaks near the wall~\cite{kline1967structure}. These streaks are amplified and distorted by nonlinear interactions and secondary instabilities, eventually leading to bursting events, ejections, sweeps, and enhanced Reynolds stresses~\cite{schoppa2002coherent}. Their breakdown regenerates vortical motions, thereby closing the feedback loop of the near-wall cycle~\cite{hamilton1995regeneration,waleffe1997self}. Thus, the coherent structures involved in the cycle should not be regarded as isolated entities, but as dynamically connected manifestations of a recurrent process through which turbulence redistributes momentum and energy in the near-wall region. This classical picture, however, does not determine which individual events within the cycle are most relevant to skin-friction generation. The VS framework is designed to address this gap.

The SHAP structures provide a data-driven view of the friction-generation cycle and address an open question from prior work. \citet{hoyas2025deep} showed that sweeps are the dominant classical structures for wall-shear prediction, but with substantial variability: many sweeps carry high enstrophy while contributing little to $\phi^{c_f}$, whereas a minority of small sweeps accounts for a disproportionate share of the friction signal. The present results make it possible to identify these high-importance events. They correspond to the wall-shear-relevant component of VS pairs, appearing downstream of velocity-relevant regions associated with high-momentum motions approaching the wall. The VS framework thus provides the missing selection criterion: not geometry or enstrophy alone, but the spatial and temporal coupling between a skin-friction-relevant event and an upstream velocity-relevant motion.

This interpretation is shown schematically in \autoref{fig:cf_cycle_shap}, which represents the VS cycle as a data-driven description of how coherent motions in the buffer layer are associated with skin-friction generation. The cycle begins with a velocity-based SHAP structure associated with a high-momentum sweep-like motion approaching the wall. As this motion is advected downstream, it is followed by a localized skin-friction-relevant response. This response is captured by the friction-based SHAP structures, which appear preferentially in the wake of the velocity-based structures and identify the regions of the velocity field most informative for the reconstruction of the local skin-friction distribution. The statistics reported in \autoref{fig:cf_cycle_shap} are consistent with the dominant role of sweeps in wall-shear generation~\cite{hoyas2025deep}. In that study, the sweeps were shown to be the most influential classical structures, with approximately twice the importance per unit volume for reconstructing wall friction compared to ejections and streaks. In contrast, for the velocity reconstruction task, sweeps, ejections, and streaks exhibit comparable importance per unit volume.

The paired organization of these structures is confirmed by the statistical analysis of their centers of gravity. \autoref{fig:cg_pairs_shap} shows the relative position of the velocity- and friction-based SHAP structures with respect to the center of gravity of their overlap. For most pairs, the center of gravity of the velocity-based structure is located upstream of that of the friction-based structure (73\% of structures), with a typical separation of approximately $10\%$ of the total pair length (dark curves of the joint probability density function). This distance exhibits a mean of $4$ wall units and a standard deviation of $17$ wall units. The few events that do not follow this trend mostly correspond to large connected clusters in which several structures are merged. This spatial ordering becomes even clearer when the probability density function is weighted by the mean SHAP value associated with friction reconstruction in the overlapping region. The most relevant skin-friction events are therefore those in which the friction-based structure appears in the wake of the velocity-based structure. These pairs of objects constitute the VS structures.

Beyond this spatial organization, \autoref{fig:cf_cycle_shap} also illustrates the temporal evolution of the VS pairs. As a VS pair is advected over the wall, the velocity-based component persists while the associated frictional response extends downstream, leading to a preferential streamwise elongation of the paired structure. The bottom-left panel of \autoref{fig:cf_cycle_shap} quantifies this geometry through the ratios between the three principal lengths, $\lambda_1$, $\lambda_2$, and $\lambda_3$, ordered from largest to smallest. The dominance of $\lambda_1$, together with its preferential alignment in the streamwise direction, shows that VS pairs are primarily elongated along the flow.

In some cases, the velocity-based component moves away from the wall as the frictional response weakens. The elongated paired structure may then fragment into smaller VS events. This process occurs in $6.6\%$ of the cases, whereas most VS pairs, $93.4\%$, dissipate before dividing. The resulting fragments can persist and interact with new velocity-relevant events, providing a possible pathway for the cycle to continue. The lifetime of the structures exhibits an exponential decay. The mean lifetime of the VS pairs is $22$ viscous time units, with a maximum observed lifetime of $155$, as shown in \autoref{fig:cf_cycle_shap}. To obtain these results, each structure is matched to the closest center of gravity of the VS pair for the preceding snapshot. The matching is constrained by a maximum allowable distance, defined as the minimum between 50\% of the structure diagonal and ${2\pi}/{5}$ wall units. The division of a structure is detected when multiple distinct structures are mapped back to the same parent VS pair. Overall, the VS cycle provides a data-driven framework for describing how individual coherent events organize the formation, evolution, and decay of wall-shear signatures in the near-wall region.

In conclusion, we have introduced VS structures as a data-driven description of skin-friction generation in turbulent channel flow. These structures are extracted from SHAP attribution fields of neural-network models trained to reconstruct the future velocity and local skin-friction fields, rather than from prescribed geometrical, kinematic, or intensity-based criteria. The resulting attribution fields organize into paired events: an upstream velocity-relevant component associated with high-momentum transport, and a downstream skin-friction-relevant component marking the corresponding wall-shear response. Their temporal evolution defines a near-wall friction cycle in which structures elongate, weaken, and either dissipate or fragment into new events. Thus, the VS framework shifts the definition of near-wall coherent structure from what the flow looks like to what it does to the wall shear, and suggests that explainable deep learning can provide a physically interpretable tool for identifying the structures most relevant to drag generation. Future work at higher $Re_\tau$ will be needed to assess the universality of the VS cycle across regimes with progressively increasing scale separation.

\section*{Acknowledgements} 
SH acknowledges support by IN/AEI/10.13039/501100011033 and by ERDF, “A way of making Europe”, under project PID2024-162480OB-I00 and CIPROM/2024/35 by GVA. RV acknowledges financial support by the University of Michigan. The authors used Claude (Anthropic, version Sonnet 4.5) to assist with language editing and manuscript preparation. The authors take full responsibility for the content of this manuscript.

\section*{Credit statement}
\textbf{Andres Cremades}: Writing – original draft, Writing – review \& editing, Investigation, Formal analysis, Conceptualization, Visualization
\textbf{Sergio Hoyas}: Investigation, Data curation, Writing – original draft,  Writing – review \& editing, Resources, Funding acquisition.   \textbf{Ricardo Vinuesa}: Conceptualization, Investigation, Writing – review \& editing, Resources,  Supervision, Funding acquisition.

\bibliography{cf_bib}

\clearpage
\newpage

\appendix
\section*{End matter}

\begin{figure}[b]
    \centering
    \includegraphics[width=1\linewidth]{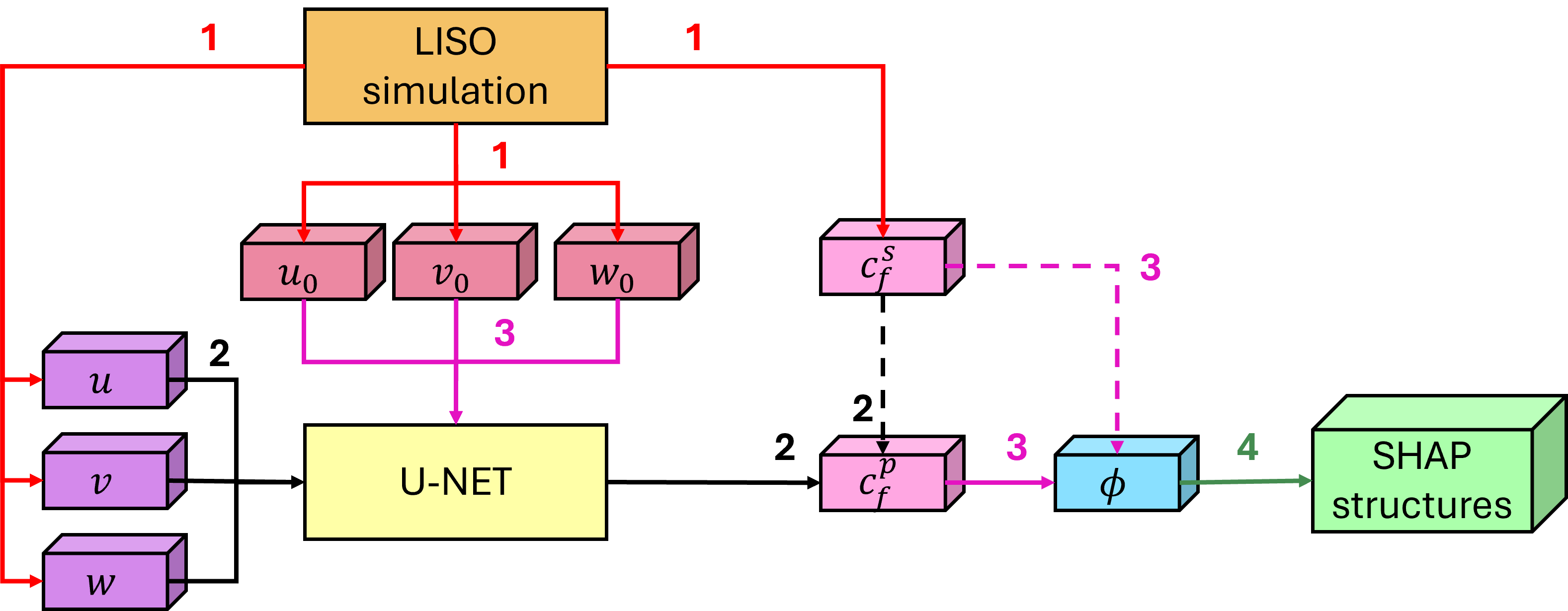}
    \caption{Visualization of workflow of the article.}
    \label{fig:workflow}
\end{figure}

The workflow, illustrated in \autoref{fig:workflow}, consists of four stages. First, a direct numerical simulation (DNS) of turbulent channel flow is used to generate the velocity fields, compute the flow statistics, and obtain the local skin-friction coefficient at the wall. The Navier--Stokes equations are integrated using the LISO code~\citep{hoyas2024turbulent}, which is based on the spectral formulation of~\citet{kim87}. The solver employs a seven-point compact finite-difference scheme in the wall-normal direction, with fourth-order consistency and extended spectral-like resolution~\citep{lele1992}, together with a third-order semi-implicit Runge--Kutta scheme for time advancement~\citep{spalart1991}. After the simulation, each instantaneous velocity field is paired with the local skin-friction distribution at a later time, $\mathbf{u}(x,y,z,t) \longrightarrow c_f(x,z,t+\Delta t)$, where $\mathbf{u}=(u,v,w)$ is the three-dimensional velocity field and $c_f$ is the local skin-friction coefficient, proportional to the wall-shear stress.

Second, a deep-learning model based on a U-Net architecture is trained to reconstruct the future local skin-friction field from the three-dimensional velocity field. The architecture follows the U-Net approach of~\citet{ronneberger2015u} and is adapted from the model used in~\citet{cremades2024}. In the present application, the input is a three-dimensional velocity field, whereas the target is a two-dimensional field defined at the wall. Therefore, the network includes an additional wall-normal compression block that transfers information from the three-dimensional encoder to the two-dimensional decoder.

\begin{figure*}
    \centering
    \includegraphics[width=0.8\linewidth]{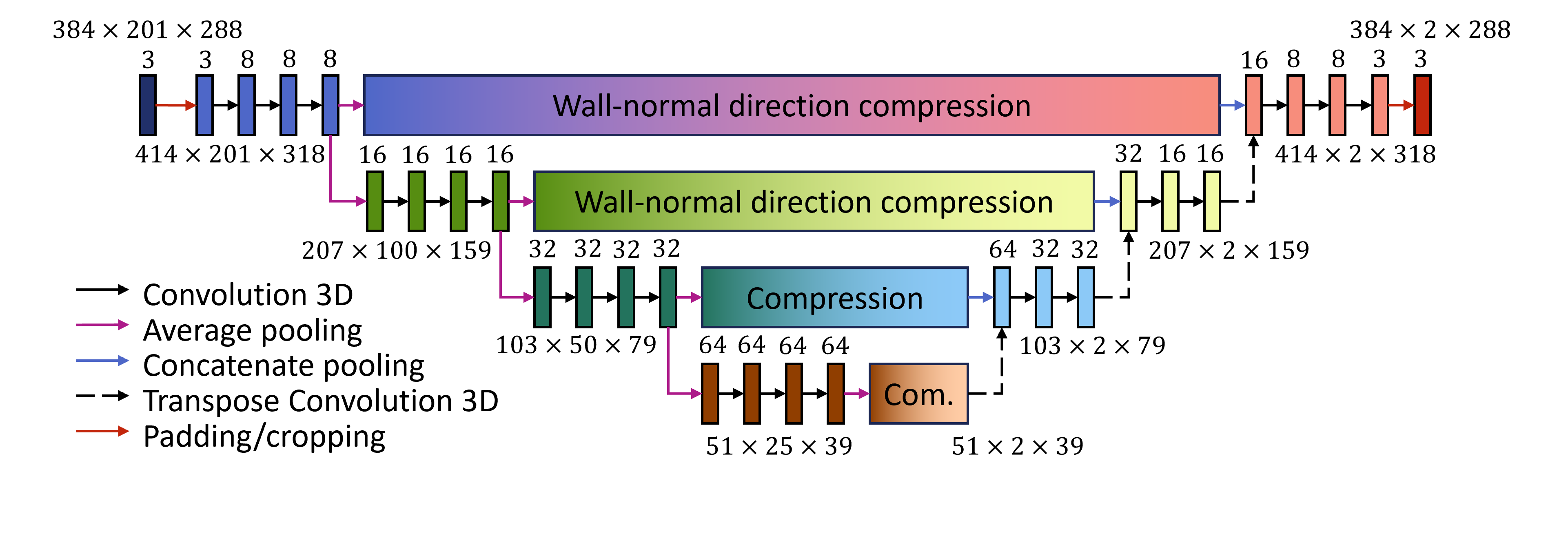}
    \caption{Architecture of the U-Net.}
    \label{fig:unet}
\end{figure*}

The U-Net comprises a total of four levels. Before entering the U-Net, the original data is expanded using a padding of $15$ grid points, which exploits the periodicity of the domain in order to avoid edge problems. The first level of the encoder applies 3 convolutional layers with kernel $3\times 3\times 3$  and fields of size $414 \times 201 \times 318$, while the decoder applies 3 convolutional layers (kernel $3\times 3\times 3$)  with a domain size $414 \times 2 \times 318$. The compression of the wall-normal direction requires a block with 5 mean pooling layers of size $1\times 4\times 1$ and 4 three-dimensional convolutions of size  $1\times 40\times 1$ and one of size $1\times 29\times 1$. For the second and third layers, the encoder uses 3 convolutional layers and the decoder 2, all of them with kernel $3\times 3\times 3$. In the second level, the tensors have dimensions of $207 \times 100 \times 159$ for the encoder and $207 \times 2 \times 159$ for the decoder. The connection between them uses 5 mean pooling layers of size $1\times 4\times 1$ and 4 three-dimensional convolutions of size  $1\times 20\times 1$ and one of size $1\times 8\times 1$. The third layer uses fields of size $103 \times 50 \times 79$ in the encoder and $103 \times 2 \times 79$ in the decoder. The compression of the wall-normal direction is calculated using 4 mean pooling layers of size $1\times 4\times 1$, 3 convolutions of kernel $1\times 10\times 1$ and 2 convolutions of kernel $1\times 1\times 1$. The fourth layer is only calculated for the encoder, using 3 convolutions with fields of size $51 \times 25 \times 39$. The compression before the decoder of the previous level requires 3 mean pooling layers of size $1\times 4\times 1$ and 3 convolutions of size $1\times 5\times 1$ and one of size $1\times 3\times 1$. A schematic representation of the U-Net is presented in \autoref{fig:unet}. The deep learning model is trained until the prediction error is below 1\%.

Third, SHAP values are computed to identify which regions of the input velocity field are most relevant to the reconstruction. Since the network maps a three-dimensional velocity field onto a two-dimensional wall quantity, the model output must first be reduced to a scalar quantity for attribution. For each trained model, we define a scalar reconstruction functional $F^k(\mathbf{u})$, where $k$ denotes the reconstructed quantity. In the present work, $F^k$ is taken as the mean-squared reconstruction error between the predicted and reference fields. The SHAP explanation is then written as an additive model,
\begin{equation}
\label{eq:mse_shap}
F^k(\mathbf{u}) \approx G(z_{ji}) =
\phi_0^k + \sum_{j\in(u,v,w)}\sum_{i=0}^{N}\phi_{ji}^k z_{ji},
\end{equation}
where $z_{ji}$ denotes the presence of the $i$-th input feature associated with the velocity component $j$, $\phi_0^k$ is the baseline contribution, and $\phi_{ji}^k$ is the attribution assigned to that feature.

The attribution fields are obtained using the GradientSHAP algorithm~\citep{erion2021}, which combines ideas from Shapley values~\citep{lundberg2017unified} and Integrated Gradients~\citep{sundararajan2017}. In this approach, the SHAP value of feature $i$ is approximated as:
\begin{equation}
\label{eq:expected_gradient}
\phi_i^k(\mathbf{u}) =
\mathbb{E}_{\mathbf{u}_0,\alpha}
\left[
\left(\mathbf{u}_i-\mathbf{u}_{0_i}\right)
\frac{\partial F^k\left(\mathbf{u}_0+\alpha(\mathbf{u}-\mathbf{u}_0)\right)}
{\partial \mathbf{u}_i}
\right],
\end{equation}
where $\mathbf{u}_0$ is a reference field and $\alpha\in[0,1]$ interpolates between the reference and the input field. The resulting attribution fields quantify the contribution of each input location and velocity component to the reconstruction functional. Thus, $\phi^u$ identifies regions relevant to the reconstruction of the future velocity field, while $\phi^{c_f}$ identifies regions relevant to the reconstruction of the future local skin-friction field.

The SHAP structures are identified using the percolation criteria given by 

\begin{eqnarray*}
\label{eq:percolation_SHAP1}
\sqrt{\left(\phi_{u}^{k}\left(x,y,z,t\right)\right)^2+\left(\phi_{v}^{k}\left(x,y,z,t\right)\right)^2+\left(\phi_{w}^{k}\left(x,y,z,t\right)\right)^2} > \\
H \sqrt{\overline{\left(\phi_u^{k}\left(y\right)\right)^2}+\overline{\left(\phi_v^{k}\left(y\right)\right)^2}+\overline{\left(\phi_w^{k}\left(y\right)\right)^2}},
\end{eqnarray*}   

\noindent where $H$ is the thresholding parameter and the overbar denotes the averaging operation used to define the wall-normal RMS profile. The percolation index has a value of $2$ for the velocity-based and $1.6$ for the friction-based SHAP structures. Connected regions of the thresholded field are then extracted as individual SHAP structures using the percolation method of~\citet{Lozano2012}. The velocity-based structures are obtained from $\phi^u=(\phi_u^u,\phi_v^u,\phi_w^u)$, whereas the skin-friction-based structures are obtained from $\phi^{c_f}=(\phi_u^{c_f},\phi_v^{c_f},\phi_w^{c_f})$.

\begin{figure} \centering \includegraphics[width=1\linewidth]{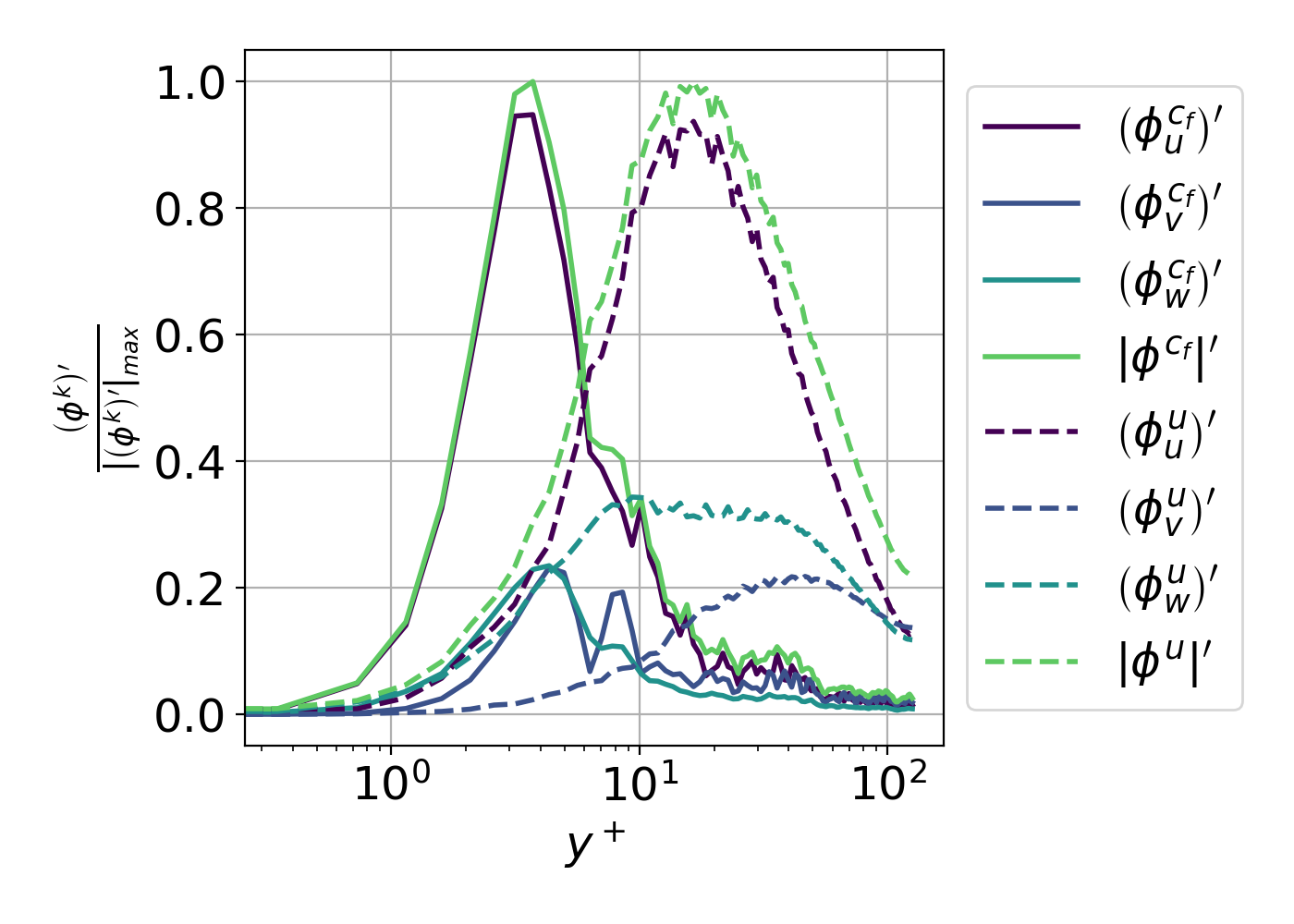} \caption{Intensity of the friction- and the velocity-based SHAP values. } \label{fig:cfshap_statistics} \end{figure}

Finally, the wall-friction cycle is defined for a wall-normal distance below $40$ wall units. This threshold is motivated by the wall-normal distribution of the SHAP intensity. \autoref{fig:cfshap_statistics} shows the intensity, $\left(\phi^k\right)'=\sqrt{\overline{\left(\phi^k\right)^2}}$, of friction-based SHAP and the velocity-based SHAP. While the former shows a peak at approximately $4$ wall units, where dissipation dominates the problem~\citep{mansour1988reynolds}, the latter peaks at $15$ wall units, in the maximum production region. Moreover, above $40$ wall units, the friction-based SHAP is negligible, and the velocity-based SHAP is half as intense.

\end{document}